\documentclass[epj]{svjour}
\newif\ifpdf            % is this pdflatex or just latex?
\ifx\pdfoutput\undefined \pdffalse \else \pdftrue \fi
\ifpdf % am spitting out pdf
\usepackage[pdftex]{graphicx}
\DeclareGraphicsExtensions{.jpg,.pdf,.png,.mps}
\else % am spitting out dvi
\usepackage[dvips]{graphicx}
\DeclareGraphicsExtensions{.eps,.ps}
\fi
\usepackage{bm}
\def\vep{\varepsilon}
\def\matr#1{\underline{\underline{{\bm{#1}}}}}

\def\lz{\ell_{\parallel}}
\def\lp{\ell_{\bot}}
\def\bea{\begin{eqnarray}}
\def\eea{\end{eqnarray}}

\begin{document}
\title{Linear hydrodynamics and viscoelasticity of nematic
elastomers}
\author{ E.M. Terentjev  \and M. Warner}
\institute{ Cavendish Laboratory, University of Cambridge,
Madingley Road, Cambridge CB3 0HE, U.K. }
\date{\today}

\abstract{We develop a continuum theory of linear viscoelastic
response in oriented monodomain nematic elastomers. The expression
for dissipation function is analogous to the Leslie-Ericksen
version of anisotropic nematic viscosity; we propose the relations
between the anisotropic rubber moduli and new viscous
coefficients. A new dimensionless number is introduced, which
describes the relative magnitude of viscous and rubber-elastic
torques. In an elastic medium with an independently mobile
internal degree of freedom, the nematic director with its own
relaxation dynamics, the model shows a dramatic decrease in the
dynamic modulus in certain deformation geometries. The degree to
which the storage modulus does not altogether drop to zero is
shown to be both dependent on frequency and to be proportional to
the semi-softness, the non-ideality of a nematic network. We
consider the most interesting geometry for the implementation of
the theory, calculating the dynamic response to an imposed simple
shear and making predictions for effective moduli and
(exceptionally high) loss factors.
} %end of abstract
\authorrunning{Terentjev  and  Warner}
\titlerunning{Viscoelasticity of nematic elastomers}
 \maketitle
\section{Introduction}  \label{intro}

The equilibrium mechanical response of liquid crystalline
elastomers can be soft or hard depending on the relation between
the imposed strains and the nematic director, in particular, if
the director is able to respond by rotating. With such unusual
equilibrium elasticity one might expect dynamical response to be
equally unusual. If the elastic forces are small, then the return
to equilibrium is driven more weakly than in conventional systems.
How does the dynamics of internal director rotation, and the
corresponding time-dependent softening of rubber-elastic response,
determine the dynamic mechanical response of a nematic rubber to a
small amplitude oscillatory shear?

In a dynamic-mechanical study of Gallani et al. \cite{hilliou96} a
stress response to an imposed oscillating shear has been studied.
Although a polydomain elastomer has been examined, the authors
reached a conclusion that the response ``is insensitive to the
isotropic-nematic transition'' and only obtained a non-trivial
result in the smectic-A phase. Subsequent studies of aligned
monodomain elastomers \cite{weileppEPL} also did not find any
unusual effect in the nematic phase and went on to investigate the
mechanical effects in the smectic-A phase. The elastic properties
of smectic rubbers are very interesting on their own, with a
number of spectacular effects even in equilibrium, see
\cite{twA,weileppA,nishikawa,osborne}. However, here we would like
to address a more basic and physically clear case of nematic
elastomers, that is, rubbery networks with an aligned uniaxial
anisotropy of their polymer strands.

Some of the reasons that no exceptional effects were found in the
nematic phase by \cite{hilliou96,weileppEPL} could be that (i) the
nematic region for the materials studied was only $\sim 7^{\rm o}$
and smectic pre-transitional effects were important, (ii) the
authors aimed to plot the whole range of dynamic modulus,
including very high glassy values at low temperatures, thus
masking a subtle nematic region and (iii) they did not study low
enough frequencies. More recently  \cite{letter}, it has been
demonstrated, both theoretically and experimentally, that a
dramatic reduction of storage modulus $G'$ and the associated
increase in the loss factor $\tan \delta$ should be expected and
indeed found in monodomain nematic elastomers sheared in certain
geometries (cf. Fig.~\ref{shears}). This effect allows one to
directly probe the basic equilibrium properties of nematic rubbers
and also access the new kinetic parameters -- viscous coefficients
and relaxation times.

In order to study the dynamics of mechanical response, one needs
to model viscoelastic properties, that is, describe viscous
dissipation in a system moving towards its equilibrium. At
present, there is no microscopic model that would even
approximately describe the dynamics of anisotropic rubbery
networks. However, as in the continuum theory of liquid crystals,
much progress can be made on a phenomenological level, using the
symmetries of variables contributing to the physical effects.

The continuum dynamics of anisotropic fluids describes macroscopic
phenomena on the level of coarse-graining analogous to that of the
nematic Frank elastic free energy. By movements of a ``liquid
particle'' one understands bodily  translations and also changes
of local director orientation of a physically infinitesimal
volume, generally of the correlation size $\xi_{\rm N} \sim 10$nm,
including many molecules in thermal equilibrium with the
reservoir, and characterised by the local nematic tensor order
parameter $Q_{\rm ij}(T)=Q(n_{\rm i}n_{\rm j} -
\frac{1}{3}\delta_{\rm ij})$. Two physical fields describe the
state of motion of nematic liquid: the local variation of director
orientation $\delta \bm{n}(\bm{r},t)=\bm{n}-\bm{n}_{\rm 0}$ with
respect to the equilibrium $\bm{n}_{\rm 0}$ and the fluid velocity
$\bm{v}(\bm{r},t)$, the time derivative of the corresponding
elastic displacement $\bm{u}(\bm{r},t)$ in the description of
elastic response. The original derivation of hydrodynamic
equations for the nematic liquid, due to Ericksen and Leslie  is
presented in some detail in key monographs on liquid crystals,
e.g. \cite{degenbook}. Here we only discuss the matters relevant
for the basic description of nematic elastomers, omitting many
fine and subtle points of this complicated subject, e.g.
\cite{TCL}. We develop a formalism describing the linear
viscoelasticity of an elastic medium with an independently
relaxing director degree of freedom and make a connection between
the (anisotropic) rubber-elastic and viscous coefficients in such
a system. The nematic dynamics coupled to the underlying rubber
elasticity leads to an expression for the entropy production
density, the Rayleigh dissipation function in the Leslie-Ericksen
form, determined by the fluxes: $\matr{\dot{\vep}}$ - the
symmetric strain rate and $\frac{d}{dt}(\bm{\Omega}- \bm{\omega})$
- the rate of relative rotation between the nematic director and
the elastic matrix:
\begin{eqnarray}
T\dot{s} &=& A_1 (\bm{n}\cdot {\matr{\dot{\vep}}} \cdot \bm{n})^2
+ 2A_4 [\bm{n} \times {\matr{\dot{\vep}}} \times \bm{n}]^2 + 4A_5
([\bm{n}\times {\matr{\dot{\vep}}} \cdot \bm{n}])^2 \nonumber
\\ &+& \textstyle{\frac{1}{2}}\gamma_{\rm 1} [\bm{n} \times
{\textstyle{ \frac{d}{dt}}}(\bm{\Omega}-\bm{\omega})]^2 +
\gamma_{\rm 2} \, \bm{n}\cdot {\matr{\dot{\vep}}} \cdot [ \bm{n}
\times {\textstyle{ \frac{d}{dt}}}(\bm{\Omega}-\bm{\omega})] \ ,
\nonumber
\end{eqnarray}
where the new linear viscous coefficients, $A_{\rm i}$, are linear
combinations of the classical Leslie coefficients, which in
elastomers can take extraordinarily high values \cite{letter}
compared with simple nematics.  $\gamma_{\rm 1}$ and $\gamma_{\rm
2}$ describe the same kind of losses involving the rotating
director as they do in a conventional nematic liquid. We find a
direct proportionality between the symmetry-related elastic and
viscous constants, connected by the corresponding relaxation times
for each deformation mode. One of the main results of our analysis
is a new dimensionless number, which we call ${\sf Ne} = \eta v/(L
\mu)$, where $\eta$ is a viscosity, $\mu$ is a rubber shear
modulus and $L$ is a characteristic length. ${\sf Ne}$ determines
the local balance between viscous and elastic torques in the
material, in analogy with the Ericksen number $Er$
\cite{degenbook}.

An essential part of description developed in this paper is the
separation of time scales. We argue, on the basis of comparison
with available experiments and qualitative arguments, that the
characteristic relaxation time of director motion is much greater
than the time scale of mechanical relaxation in an ideal polymer
network. Accordingly, we only consider the low-frequency limit of
rubber viscoelastic response, reflected in essentially linear
frequency dependence of complex modulus $G^* = G_{\rm o} + {\rm i}
\omega \eta$. In contrast, the much slower director relaxation
allows us to examine both its low- and high-frequency regimes.

This paper is organised as following. The next Section, continuing
the introduction into the subject, briefly sketches the
fundamentals of equilibrium nematic rubber elasticity, focusing on
the linear continuum description rather than the full molecular
theory valid to high deformations. Following this, in
Section~\ref{hydro} we revisit the Leslie-Ericksen formulation of
the continuum dissipation function $T\dot{s}$ and show that, as
would be expected by symmetry, it is identical to the elastic
energy density only with strains substituted by strain rates. We
then discuss the fundamentals of nematic rubber viscoelasticity.
In Section~4 the practical calculation of dynamic moduli
$G^*(\omega)$ is presented for three principal shear geometries,
as an example of general theory implementation. The results
indicate which combinations of elastic and viscous coefficients
are relevant for the response and, therefore, directly measurable
by experiment. We then conclude by discussing the role of soft
elasticity in slowing the mechanical relaxation and make contact
with new experimental data.

\section{Nematic rubber elasticity} \label{eqsec}

Equilibrium elastic properties of monodomain nematic rubbers are
well-studied, both theoretically and experimentally and are
described at some length in review articles \cite{lce,review1}. A
full molecular theory of ideal nematic networks gives the elastic
free energy density
 \begin{equation}
F={\textstyle{\frac{1}{2}}} c_{\rm x} k_{\rm B}T \ {\sf
Tr}(\matr{\lambda}^T \cdot \matr{\ell}_\theta^{-1} \cdot
\matr{\lambda} \cdot \matr{\ell}_0) \ + {\textstyle{\frac{1}{2}}}
\widetilde{B} \big( {\sf Det} [\matr{\lambda}] -1 \big)^2,
\label{F0}
 \end{equation}
where $\matr{\lambda}$ is a Cauchy strain tensor, $\lambda_{\rm
ik}=\partial R_{\rm i}/\partial R^0_{\rm k}$. In a nematic
elastomer network, $\matr{\ell}$ are the uniaxial matrices of
chain step-lengths before ($0$) and after the director $\bm{n}$
has rotated by a certain angle $\theta$ during the deformation:
$\ell_{ij} = \lp \delta_{ij} + [\lz-\lp]\, n_i \, n_j$. The last
term in (\ref{F0}), the bulk-modulus contribution independent of
the configurational entropy of polymer chains, is determined by
molecular forces resisting the compression of a molecular liquid,
$\widetilde{B} \sim 10^{9} \, \hbox{J/m}^3$, much greater than the
typical value of rubber modulus $\mu \simeq c_{\rm x} k_{\rm B}T$,
with $c_{\rm x}$ the effective crosslinking density, giving $\mu
\sim 10^{5} \, \hbox{J/m}^3$. This large bulk energy penalty
constrains the value of the strain determinant, ${\sf Det} [
\matr{\lambda}] \approx 1$ (which, in other words, means that the
material is physically incompressible when subjected to all
strains except hydrostatic compressions and dilations).
Eq.~(\ref{F0}) is only limited by the Gaussian network assumption
and is valid up to very large strains where mechanical softness,
rotational instabilities and optical effects occur - well beyond
the limits of continuum theory.

Olmsted \cite{olm}, in studying soft elasticity -- shape changes
in nematic elastomers without energy cost, has derived the
small-deformation continuum limit of the full expression
(\ref{F0}), when $\lambda_{\rm ij}= \delta_{\rm ij} +
\partial_{\rm i} u_{\rm j}$ and $| \nabla \bm{u} | \ll 1$. We shall use
a slightly different notation, more suitable for the
linear-elastic description because it better complies with the
standard textbook formalism of uniaxial elasticity, e.g.
\cite{landau}. The small-deformation limit of (\ref{F0}) is
\begin{eqnarray}
F &=& C_1 (\bm{n}\cdot \tilde{\matr{\vep}} \cdot \bm{n})^2 + 2C_2
{\rm Tr} [\matr{\vep}](\bm{n}\cdot \tilde{\matr{\vep}} \cdot
\bm{n})+ C_3 \left( {\rm Tr}[\matr{\vep}] \right)^2 \nonumber  \\
&+& 2C_4 [\bm{n} \times \tilde{\matr{\vep}} \times \bm{n}]^2 +
4C_5 ([\bm{n}\times \tilde{\matr{\vep}} \cdot \bm{n}])^2
\label{mainF}
\\ &+& \textstyle{\frac{1}{2}}D_1 [\bm{n} \times
(\bm{\Omega}-\bm{\omega})]^2 + D_2 \, \bm{n}\cdot
\tilde{\matr{\vep}} \cdot [ \bm{n} \times
(\bm{\Omega}-\bm{\omega})]. \nonumber
\end{eqnarray}
where $\bm{n}$ is the undistorted nematic director.
$\tilde{\vep}_{\rm ik}=\vep_{\rm ik}-\frac{1}{3}{\rm Tr}
[\matr{\vep}] \, \delta_{\rm ik}$ is the traceless part of linear
symmetric strain $\vep_{\rm ik} = \frac{1}{2}(\partial_{\rm k}
{u}_{\rm i} + \partial_{\rm i} {u}_{\rm k})$, which is the only
variable of classical continuum elasticity \cite{landau}. In a
system with an internal orientational degree of freedom, the
nematic director with its own relaxation and dynamics, the
antisymmetric part of strain expressed by the local rotation
vector $\bm{\Omega}=\frac{1}{2} {\rm curl} \, \bm{u}$ may
contribute to the physical properties. Analogously, the small
vector $\bm{\omega} $ is a convenient measure of director
rotations, $\bm{\omega} = [\bm{n}\times \delta \bm{n}]$. In fact,
it is the relative rotation, the difference $\bm{n} \times
(\bm{\Omega}-\bm{\omega})$, that causes the elastic response and a
number of effects unique to nematic elastomers \cite{degenD}.

One expects that in a rubber or dense polymer melt the bulk
modulus $C_3$ is very large, $C_3 \sim \widetilde{B}$. We,
therefore, shall only consider deformations with no bulk
compression: ${\rm Tr} [\matr{\vep}]=0$. In general, all other
constants in the expression (\ref{mainF}) are of the same order of
magnitude, similar to the rubber modulus $\mu$. The molecular
model of ideal nematic elastomer \cite{review1} gives specific
forms for these constants:
\begin{eqnarray}
C_1 &=& 2 C_4 = c_{\rm x}k_{\rm B}T, \ \ \ C_5={\textstyle
\frac{1}{8}}c_{\rm x}k_{\rm B}T \frac{(r+1)^2}{r},  \nonumber
\\ D_1 &=& c_{\rm x}k_{\rm B}T \frac{(r-1)^2}{r}, \ \ \ D_2=c_{\rm
x} k_{\rm B}T \frac{1-r^2}{r} , \label{params}
\end{eqnarray}
in which case the condition for ideal soft elasticity holds,
\begin{equation}
C^R_5 = C_5-\frac{D_2^2}{8 D_1}=0 .\label{renorm}
 \end{equation}
Model expressions for elastic constants (\ref{params}) depend,
apart from the universal rubber-elastic energy scale $\mu$, on a
single parameter $r$. In the molecular model (\ref{F0}) of  ideal
nematic polymer networks one finds that $r=\lz/\lp$, the ratio of
the principal step lengths of the anisotropic polymer backbone (or
equivalently $ r=(R_{\|}/R_{\bot})^2$ in terms of the  principal
values of gyration radii tensor). In non-ideal elastomers, this
parameter is more complex, determined by a number of other
factors, for instance when there are fluctuations in composition,
see the Appendix. Nevertheless, it has to be a function of nematic
order parameter $Q(T)$, satisfying a linear limit $r\approx 1 +
\beta \, Q$, at least at small $Q$. In the isotropic phase, at
$Q=0$ and $r=1$, the elastic constants (\ref{params}) become, as
expected: $C_1=2C_4=2C_5=\mu$, $D_1=D_2=0$ and the elastic energy
(\ref{mainF}) reduces to a standard Lam\'e expression. See the
Appendix for a discussion of non-ideality, that is where $C^R_5
\ne 0$.

\begin{figure}
  \centering
%\resizebox{0.37\textwidth}{!}{\includegraphics{geom.eps}}
\resizebox{0.37\textwidth}{!}{\includegraphics{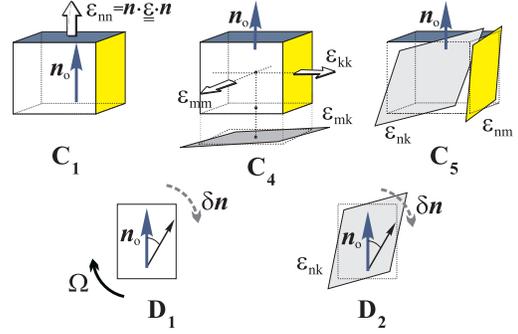}}
\caption[]{The symmetries of the shears and rotations possible in
a uniaxial solid at constant volume and with an internal
orientational degree of freedom.  The modes are labeled with the
appropriate elastic constant and symmetric shear or rotation.
$C_4$ also shows a section perpendicular to the director to
illustrate shears $\varepsilon_{\rm mk}$ not involving $\bm{n}$
(the directions $\bm{m}$ and $\bm{k}$ are perpendicular to
$\bm{n}$). In the two coupling terms, $D_1$ and $D_2$, the small
director rotation $|\bm{\omega}| \approx |\delta \bm{n}|$. }
\label{geom}
\end{figure}

In an incompressible material, all deformations are essentially
shears, albeit sometimes seen in a rotated frame. Figure
\ref{geom} shows the character of these shears in the
eq.~(\ref{mainF}) and their corresponding elastic constants. $C_1$
``lives" along the director, $C_4$ in the plane perpendicular to
$\bm{n}_0$ in which properties are isotropic ($\varepsilon_{\rm
kk}$, $\varepsilon_{\rm mm}$ and $\varepsilon_{\rm mk}$ are of the
same status since the solid is uniaxial).  Most interesting are
$C_5$, involving shears such as $\varepsilon_{\rm nk}$, which span
directions parallel and perpendicular to the initial director and
which can induce it to rotate; $D_1$ which resists director
rotations with respect to the solid matrix; and $D_2$ where the
rotation is coupled to the symmetric shears also involved in
$C_5$. The inter-relation between these three processes is what
gives rise to the effect of soft elasticity -- here shape change
of the type $\varepsilon_{\rm nk}$ without energy cost.  If the
director is allowed to react ideally and is regarded as a slave
variable, then the renormalised resistance to deformations
$\varepsilon_{\rm nk}$, that is $C^R_5$, vanishes as we see in
eq.~(\ref{renorm}). We shall see, in Section~4 for simple shear
geometries, the mechanism by which this is achieved -- at least at
harmonic order (since these simple geometries do not allow the
full shear and extensional freedom required for softness at all
amplitudes).

\section{Nematohydrodynamics of elastic solids}\label{hydro}
When nematic elastomers are strained, there is director rotation
along with stresses -- the latter  yielding both body forces and
torques. At finite strain and rotational rates there will be
stresses of both elastic and viscous origin, the latter being
conceptually parallel to those arising in classical liquid
nematics.  We shall accordingly review classical
nematohydrodynamics, ignoring the subtleties that reside at the
heart of the subject.  There are important differences with
liquids -- if torques are induced by flow, they will be balanced
in solids, on length scales longer than the nematic penetration
depth, by torques generated by the matrix, that is, by $D_1$ and
$D_2$ elastic terms that involve the anti-symmetric part of the
strain $\matr{\lambda}$. The issue of possible director gradients
and corresponding Frank elasticity in nematic elastomers has been
extensively discussed in the literature \cite{lce,review1,chol}.
It is known that, unless there are special reasons for a director
singularity (such as in disclinations or narrow domain walls),
Frank elastic effects play a minor role in the free energy balance
and can be neglected. Instead, the rubber-elastic matrix generates
the torques, which shape the final character of the dynamics. A
new dimensionless number will be introduced to replace the
Ericksen number describing a torque balance in ordinary nematics.
There is another important difference -- in liquids, as flow
proceeds, the strain and the director rotation can increase
without limit, whereas in solids both are constrained (and are
coupled to each other) by the matrix. Elastomers are capable of
huge distortions but we shall limit ourselves to linear theory.
Nematohydrodynamics in the elastically highly non-linear regime is
a subject of another study.

\subsection{Leslie-Ericksen formulation of nematic viscosity}

The equation of motion can be written in the usual form of local
balance of forces: in vector components,
\begin{equation}
\frac{\partial}{\partial t}\big( \rho v_{\rm i} \big) = -
\frac{\partial \Pi_{\rm ki}}{\partial x_{\rm k}}  ,  \label{ns1}
\end{equation}
where $\rho$ is the density and $\Pi_{\rm ki}$ the tensor of
momentum flux density, $\Pi_{\rm ki} = \rho v_{\rm k} v_{\rm i} -
\sigma_{\rm ki} + P \, \delta_{\rm ki}$ with $P$ the local
pressure and $\sigma_{\rm ki}$ the stress tensor. In an isotropic
Newtonian liquid $\sigma_{\rm ki}$ is proportional to the
symmetric strain rate, $A_{\rm ki} =\frac{1}{2} \left(
\partial_{\rm k} v_{\rm i} + \partial_{\rm i} v_{\rm k}\right) \equiv
\dot{\vep}_{\rm ki}$. In uniaxial nematic liquid crystals, the
stress tensor depends not only on the fluid velocity gradients,
but also on the components and the gradients of the local nematic
order $Q_{\rm ki}$. Because rotations and corresponding torques
are involved if the director changes differently from the local
fluid rotation, the stress tensor is no longer symmetric and also
depends on the relative rotation combinations $\bm{N}=
\frac{d}{dt} \delta\bm{n}+ \frac{1}{2}[\bm{n} \times {\rm curl} \,
\bm{v}]$ \cite{degenbook}, which is nothing but a time derivative
of the relative-rotation combination, that is $\bm{N}=[\bm{n}
\times \frac{d}{dt}(\bm{\Omega}-\bm{\omega})]$, cf.
equation~(\ref{mainF}). As a result, Leslie writes the viscous
stress tensor
\begin{eqnarray}
\sigma_{\rm ij} &=& \alpha_{\rm 1} n_{\rm i}n_{\rm j} A_{\rm km}
n_{\rm k} n_{\rm m} +\alpha_{\rm 2} n_{\rm i}N_{\rm j} +
\alpha_{\rm 3} n_{\rm j}N_{\rm i} \label{leslie1} \\ &&  +
\alpha_{\rm 4} A_{\rm ij} + \alpha_{\rm 5} n_{\rm i} n_{\rm k}
A_{\rm kj} + \alpha_{\rm 6} n_{\rm j} n_{\rm k} A_{\rm ki}
\nonumber
\end{eqnarray}
(assuming incompressible fluid, $A_{\rm kk}= {\rm div} \,
\bm{v}=0$). Its antisymmetric part contributes to the local torque
$\bm{\Gamma} = [\bm{n} \times \bm{h}] $ (cf. \cite{degenbook})
where $\bm{h}$ is the molecular  field.  $\bm{h}$ derives also in
part from Frank and external field terms. These contributions are
balanced by the anti-symmetric terms deriving from viscous flow
processes:
\begin{eqnarray}
 h_{\rm i} &=& \frac{\partial
}{\partial x_{\rm k}} \left( \frac{\delta F}{\delta[ \nabla_{\rm
k} n_{\rm i}]} \right) - \frac{\partial F}{\partial n_{\rm i}}
=\gamma_{\rm 1} N_{\rm i} + \gamma_{\rm 2} n_{\rm k} A_{\rm ki}
 \label{leslie2}
\end{eqnarray}
with $\gamma_{\rm 1}=\alpha_{\rm 3}-\alpha_{\rm 2}$ and
$\gamma_{\rm 2}=\alpha_{\rm 3}+\alpha_{\rm 2}= \alpha_{\rm
6}-\alpha_{\rm 5}$ (the Parodi relation, a representation of the
Onsager principle of kinetic coefficient symmetry). Here the
viscous coefficients $\alpha_{\rm 1} , \, ... \, ,  \alpha_{\rm
6}$ all depend on the magnitude of nematic order parameter $Q(T)$.
One can show that in an ordinary nematic liquid crystal near the
weak first order transition, as $Q \rightarrow 0$, they should
behave as \cite{imura289}
\begin{eqnarray}
\alpha_{\rm 1} \propto Q^2, && \left( \alpha_{\rm 2}, \,
\alpha_{\rm 3}, \, \alpha_{\rm 5} , \, \alpha_{\rm 6} \right)
\propto Q  \label{viscscaling} \\
 {\rm also} && \gamma_{\rm 1} \propto Q^2 \ \
{\rm and} \ \ \gamma_{\rm 2} \propto Q . \nonumber
 \end{eqnarray}
Thus, in the isotropic phase, only one of the Leslie coefficients
survives: $\alpha_{\rm 4} \rightarrow 2\eta$. Molecular theory
\cite{kuzuu,OT} also shows that in a typical nematic liquid of
rod-like molecules these coefficients may have very different
magnitude and even sign: $ \alpha_{\rm 2}, \, \alpha_{\rm 3},  \,
\alpha_{\rm 6}$ -- negative, $ \alpha_{\rm 4}, \, \alpha_{\rm 5}$
-- positive; $( |\alpha_{\rm 2}| \sim \alpha_{\rm 5} ) \gg |
\alpha_{\rm 3}|, \, |\alpha_{\rm 6}|$; $\alpha_{\rm 1}$ is
generally small and may be positive or negative in different
materials; far from the nematic transition (at $Q \rightarrow 1$)
the ``isotropic coefficient'' $ \alpha_{\rm 4} \sim |\alpha_{\rm
2}| \sim \alpha_{\rm 5}$. De Gennes assembles the experimental
values for Leslie coefficients in MBBA at $25^{\rm o}C$
\cite{degenbook}, while the monograph by de Jeu \cite{dejeubook}
provides the values for another classical nematic, PAA at
$122^{\rm o}C$: in $10^{-3}$~Pa.s
\begin{eqnarray}
\begin{array}{rcccccccc}
& \alpha_{\rm 1}  & \alpha_{\rm 2}  & \alpha_{\rm 3}  &
\alpha_{\rm 4} & \alpha_{\rm 5} &  \alpha_{\rm 6} & \gamma_{\rm 1}
& \gamma_{\rm 2} \cr {\rm MBBA}: \ \  & 6.5  & -77.5 & -1.2 \ &
83.2 \ & 46.3 & -34.4 \ & 76.3 & -78.7 \cr
 {\rm PAA}: \ \ & 4 & -6.9 & -0.2 & 6.8 & 5 & -2 &
 6.8 & -7.1  \end{array}  \nonumber
\end{eqnarray}

\subsection{Nematic elastomers}
We now develop Leslie-Erickson theory for solid nematics by a
direct analogy.  The nematic dynamics is coupled to the underlying
anisotropic elasticity described in its linear limit by
eq.~(\ref{mainF}) and sketched in Fig.~\ref{geom}.  It is thus
natural to develop the dynamics using the same symmetry-grouping
of terms as in (\ref{mainF}), rather than as in (\ref{leslie1}) as
is usual for classical nematics.

Two differential equations, (\ref{ns1}) with the viscous stress
given by eqs.~(\ref{leslie1}), and (\ref{leslie2}), form the
complete set describing the linear viscous effects in the nematic
fluid. Neglecting the effects of heat convection, the total energy
dissipation (the entropy production) in such anisotropic medium is
expressed by the volume integral of the conjugate forces and
fluxes:
\begin{eqnarray}
T\dot{S} &=& \int dV \left( \sigma_{\rm ij} A_{\rm ij} + (\bm{h}
\cdot \dot{\bm{n}}) - (\bm{\Gamma} \cdot \bm{\Omega}) \right)
\nonumber \\ &\equiv & \int dV \left( \sigma_{\rm ij}
\dot{\vep}_{\rm ij} + (\bm{h} \cdot \bm{N} ) \right) .
\label{tsdot}
\end{eqnarray}
We re-write the density of dissipation function in a form matching
the elastic energy density (\ref{mainF}):
\begin{eqnarray}
T\dot{s} &=& A_1 (\bm{n}\cdot {\matr{\dot{\vep}}} \cdot \bm{n})^2
+ 2A_4 [\bm{n} \times {\matr{\dot{\vep}}} \times \bm{n}]^2 + 4A_5
([\bm{n}\times {\matr{\dot{\vep}}} \cdot \bm{n}])^2 \nonumber
\\ && \ \ + \textstyle{\frac{1}{2}}\gamma_{\rm 1} \bm{N}^2 +
\gamma_{\rm 2} \, \bm{n}\cdot {\matr{\dot{\vep}}} \cdot \bm{N}
\label{rayleigh}
\end{eqnarray}
in the fully incompressible case. Thus, by differentiation of
(\ref{rayleigh}) respectively by the symmetric strain rate
$\dot{\vep_{ij}}$ and by $N_i$, one obtains a representation of
the symmetric viscous stress tensor and the nematic molecular
field, contributing to the local torque, analogous to the parallel
expressions (\ref{leslie1}) and (\ref{leslie2}) for simple
nematics:
\begin{eqnarray}
\sigma_{\rm ij}\sp{s} &=& 2A_1 \left(\bm{n}\cdot
{\matr{\dot{\vep}}} \cdot \bm{n}\right)n_i n_j +
4A_4\left[\bm{n}\times  \left(\bm{n} \times {\matr{\dot{\vep}}}
\times \bm{n}\right)\times \bm{n}\right]_{ij}\nonumber \\&+&
4A_5\left( \left[\left(\bm{n}\times {\matr{\dot{\vep}}} \cdot
\bm{n}\right)\times \bm{n}\right]_i n_j + \left[\left(\bm{n}\times
{\matr{\dot{\vep}}} \cdot \bm{n}\right)\times \bm{n}\right]_j n_i
\right)\nonumber\\ &+& \textstyle{\frac{1}{2}}\gamma_{\rm
2}\left(n_iN_j +N_i n_j \right) \ ;
 \label{sigma2} \\
h_{\rm i} &=& \gamma_{\rm 1} {N}_{i} + \gamma_{\rm 2} {n}_j
{{\dot{\vep}}}_{ij}  \ , \nonumber
\end{eqnarray}
where $\sigma_{\rm ij}\sp{s}$ is the symmetric part of an
expression like (\ref{leslie1}) which emerges naturally since we
express $T\dot{s}$ in terms of symmetric and anti-symmetric
variables separately. Here the constants are linear combinations
of classical Leslie coefficients
\begin{eqnarray}
A_1 &=& \textstyle{\frac{1}{2}}(\alpha_{\rm 1} + \alpha_{\rm 4} +
\alpha_{\rm 5} + \alpha_{\rm 6}), \label{As} \\
A_4 &=& \textstyle{\frac{1}{4}} \alpha_{\rm 4}, \nonumber \\ A_5
&=& \textstyle{\frac{1}{8}}(2 \alpha_{\rm 4} + \alpha_{\rm 5} +
\alpha_{\rm 6}). \nonumber
 \end{eqnarray}
In the isotropic limit, one finds $A_1 = 2 A_4 = 2 A_5 \rightarrow
\eta$, reminiscent of the isotropic Lam\'e limit of the
corresponding elastic constants.

\subsection{Relaxation times}
The direct correspondence between the elastic constants of the
energy density $F$, eqs.~(\ref{mainF}), and the viscous
 coefficients of the dissipation function $T\dot{s}$, eq.~(\ref{rayleigh}), gives: $$
A_{\rm i}
 \rightarrow C_{\rm i}; \ \gamma_{\rm 1} \rightarrow D_{\rm 1} \
{\rm and} \ \gamma_{\rm 2} \rightarrow D_{\rm 2}. $$
 Of course,
this is not a surprise since the symmetry of an equilibrium
elastic deformation and of a viscous flow in a uniaxial continuum
is the same. Equally, there is a direct correspondence in the
dependence of the various coefficients on nematic order parameter
$Q$, see eq.~(\ref{viscscaling}) for viscosities and
eq.~(\ref{params}) for elastic constants. We can equivalently,
therefore, represent the viscous coefficients of a nematic
elastomer  by products of the corresponding rubber-elastic
constant and an appropriate relaxation time, that is,
\begin{equation}
A_{\rm i} = C_{\rm i}\tau_{\rm R}; \ \gamma_{\rm 1} =D_{\rm
1}\tau_{\rm 1}; \ \gamma_{\rm 2} = D_{\rm 2}\tau_{\rm 2}.
\label{parT}
\end{equation}
In principle, the relaxation times for the various $A_i$ will be
different, and different from the $\tau_i$. However, one expects a
significant separation between the two groups of characteristic
time scales, that is between the $\tau_{\rm R}$s and the
$\tau_i$s. The director rotation time, $\tau_{\rm 1}= \gamma_{\rm
1}/D_{\rm 1}$, has been used in a simplified, single-viscosity
analysis \cite{letter}, and then estimated experimentally as
$\tau_{\rm 1} \sim 10^{-1}-10^{-2}$s. In contrast, the
characteristic time of rubber relaxation $\tau_{\rm R}$ is
expected to be much shorter. A recent statistical theory
explicitly calculated this time from the analysis of network
strand relaxation within a tube model \cite{samjcp}, showing this
is of order of Rouse time. One could also confirm this by a
following qualitative observation: The linear viscous stress
(\ref{sigma2}) and the analogous corresponding expression for the
elastic stress could be combined, producing the effective
frequency-dependent moduli in a form $(C_{\rm i} + {\rm i} \omega
A_{\rm i})$. This is the low-frequency limit of a general complex
modulus $G^*(\omega)$, showing the rubber plateau modulus
$G'=C_{\rm i}$ and the initial rise in the loss modulus with
frequency, describing a viscous flow. In the classical picture of
polymer dynamics, e.g. \cite{DoiEd}, the next characteristic
regime is at a Rouse frequency, when the signal cannot propagate
along the polymer chain length and mechanical response is provided
by individual segments, thus causing a climb of $G^*(\omega)$
towards the glass plateau. Therefore, the characteristic time
scale $\tau_{\rm R}$ in the estimation relations (\ref{parT}) is
of the order of the Rouse time, $\tau_{\rm R} \sim
10^{-4}-10^{-6}$s.

Could there be viscous softness analogously to the elastic
softness of ideal elastomers, for instance in eq.~(\ref{renorm})?
Two related arguments suggest that this is not possible.  The
viscous combination $A_5 - \gamma_2^2/8\gamma_1$, analogous to the
renormalised elastic constant $C_5^{R}$, can be rewritten with the
benefit of timescales eq.~(\ref{parT}) as:
\begin{eqnarray} A_5 - \gamma_2^2/8\gamma_1 = \tau_{\rm R}
\left(C_5 - \frac{\tau_2^2}{\tau_{\rm R}\tau_1}\frac{D_2^2}{8
D_1}\right)\; . \label{viscsoft}\end{eqnarray} Even if we took the
rotational relaxation times equal, $\tau_1 = \tau_2$, the
translational relaxation time $\tau_{\rm R}$ would appear to upset
the possibility of renormalising (\ref{viscsoft}) to zero, even if
we had $C_5^R \rightarrow 0$. Elastic softness arises because an
anisotropic distribution of chains can be rotated undistorted and
thus at constant entropy. However, individual chains will be
distorted and there must reasonably be dissipation associated with
their flow relative to the matrix. Thus elastically soft
distortions should have associated dissipation, except perhaps
accidentally if rates of $\gamma_1$, $\gamma_2$ and $A_5$
relaxations compensate to make (\ref{viscsoft}) vanish.

\subsection{Symmetries and order parameter}\label{symmetries}

The ability to neglect complicated inertial effects in fluid
dynamics is controlled by the small magnitude of Reynolds number,
$Re = \rho \, v \, L/\eta \ll 1$, with $L$ the characteristic
length and $\eta$ the typical viscosity. In nematic liquids,
another dimensional number is introduced to characterise the
relative magnitude of hydrodynamic and Frank elastic torques
contributing to eq.~(\ref{leslie2}), where the left hand side is
$\sim K \nabla^2 n$, while the right hand side is $\sim \gamma
\nabla v$, with the rotational viscosity $\gamma \sim \eta$. Small
Ericksen number $Er = \eta \, v \, L/K \ll 1$ means that the
director orientation is mostly controlled by equilibrium elastic
free energy, while at $Er \gg 1$ the director generally follows
the local orientation provided by the flow.  In elastomers we are
concerned with the balance between flow-induced torques, again
scaling as $\sim \eta \nabla v$, and those of the rubbery matrix,
expressed by $\partial F/\partial \theta \sim D_1\theta$. This
yields a new dimensionless group characterised by the number ${\sf
Ne} = \eta v /(L D_1)$. Note that the assumed domination of
rubber-elastic over Frank effects essentially means that $D_1 \gg
K/L^2$, that is, ${\sf Ne} \ll Er$. Our analysis of an example of
simple shear deformation, in the next Section, spans the full
range of small to large ${\sf Ne}$, which we shall define more
precisely for that geometry.

We have seen in Section~2, in the discussion of
eqs.~(\ref{params}) and chain-anisotropy parameter $r$, how the
anisotropic rubber moduli depend on the underlying nematic order
parameter $Q$, usually a function of temperature or solvent
concentration. In the isotropic phase, at $Q=0$ and $r=1$, the
elastic constants return to the classical Lam\'e values
$C_1=2C_4=2C_5=\mu$, $D_1=D_2=0$. The coupling constant $D_2$
depends on the linear power of $Q$. As a result, when there is
anisotropy, $Q\neq 0$ and $r\neq 1$, the sign of the elastic
constant $D_2$ depends on whether $r>1$ (prolate order, $D_2 < 0$)
or $r<1$ (oblate order, $D_2 > 0$). The sign of director rotation
relative to the matrix, $\bm{\omega} -\bm{\Omega}$, varies
accordingly to whether chains are prolate (they align with the
extension direction associated with the shear, see Fig.~\ref{geom}
for $D_2$) or oblate (alignment along the compression axis).  In
any event, a term such as $D_2 \varepsilon_{\rm nk} \omega$,
bilinear in $\varepsilon_{\rm nk}\omega$, always reduces the
elastic energy, irrespective of the sign of $D_2$, and indeed this
is the actual mechanism whereby a shape change can ideally be
achieved with no energy cost \cite{olm,JMPS}. There is an
appealing analogy between this behaviour and the Leslie
coefficient $\gamma_{\rm 2}$ of discotic nematics. It is known
\cite{volovik,carlsson} that the different (disk- instead of
rod-like) molecular shape leads to the opposite sign of
$\gamma_{\rm 2}$, with the according consequences for the flow
alignment properties.

Ideal rubber-elastic expressions (\ref{params}) suggest that the
coupling constant $D_1$ is proportional to $(r-1)^2 \sim Q^2$.
Continuing the analogy with liquid nematic viscosity, we should
recall the estimates of Imura and Okano \cite{imura289}, but also
a following qualitative consideration. The rotational viscosity
$\gamma_{\rm 1}$ is determined by the antisymmetric part of
viscous stress tensor $\sigma'_{\rm ij}$, which cannot be
proportional to the linear power of the {\em symmetric} tensor
nematic order parameter $Q_{\rm ij}$, but at least its square.
However, there is a delicate problem arising in the analysis of
soft elasticity. The renormalised shear modulus $C^R_5 =
C_5-D_2^2/8 D_1$ should reduce to the bare $C_5 \rightarrow
\frac{1}{2}\mu$ in the isotropic phase, at $Q \rightarrow 0$.
However, the ideal values of $D_1$ and $D_2$ result in a finite
renormalisation in this limit. One must revise the conclusions of
non-ideal, semi-soft theory \cite{review1}. Eq.~(\ref{nonsoft})
and the discussion in the Appendix suggest that, in fact, $D_1$
has a additional small (semi-soft) correction $\sim Q$. This
resolves the problem of making the renormalisation $D_2^2/8 D_1$
vanish at $Q\rightarrow 0$, but raises a question about the
symmetry consideration that $\gamma_{\rm 1}$ (and $D_1$, by
parallel) cannot be linear functions of $Q_{\rm ij}$. The paradox
is safely resolved when one recognises that the semi-soft
coefficient $a_1$ in eq.~(\ref{nonsoft}) is a linear function of
$Q_f$, the order at network formation. Therefore, in fact, both
$D_1$ and $\gamma_{\rm 1}$ in nematic elastomers depend on the
bilinear combination $Q^{\{f\}}_{\rm ij}Q_{\rm ij}$, which means
$\sim | Q |$, and thus no symmetry problem arises. In an ordinary
liquid nematic there is no issue of formation order being frozen
by crosslinking, and $\gamma_{\rm 1} \sim Q^2$ as expected.

\subsection{Balance of forces and torques}
To complete the general analysis of this Section, we briefly
discuss the stresses and torques that follow from the combination
of elastic and dissipation functions. In an overdamped system,
traditionally ignoring inertial effects at low-frequency, we have
two equations of motion: the balance of forces and torques. The
first condition requires locally balancing the total symmetric
stress tensor
\begin{equation}
 \matr{\sigma}
= \matr{\sigma}_{\rm el} + \matr{\sigma}_{\rm visc}
 =
 \frac{\partial F}{\partial \matr{\vep}} +
 \frac{\partial ( T\dot{s})}{\partial \matr{\dot{\vep}}} \;\; .
\end{equation}
In the case when the relaxation time scales are separated, as in
the discussion above, the viscous stress contribution to the force
balance is minor. In contrast, the balance of torques,
$\bm{\Gamma}= [\bm{n}\times \bm{h}]=0$ in the absence of external
fields, requires the full molecular field
\begin{eqnarray}
 \bm{h}= \bm{h}_{\rm el} + \bm{h}_{\rm visc}
 = &D_1&  [\bm{n} \times (\bm{\Omega}-\bm{\omega})] + D_2 \bm{n} \cdot
 \matr{\vep} + \label{torques}\\ &+& \gamma_1 [\bm{n} \times \frac{d}{dt}(\bm{\Omega}-\bm{\omega})] + \gamma_2 \bm{n} \cdot
 \matr{\dot{\vep}} \nonumber \;.
\end{eqnarray}
Here both groups of terms are manifestly of the same order of
magnitude in the regime of frequencies characterising director
rotations. We shall see in the particular calculation of Section~4
that the condition of zero local torque allows one to obtain, for
instance, the rate of director variation $\dot{\theta}$.

In the limit of isotropic rubber $Q\rightarrow 0$ the only
relevant equation is that for the stress, which in this case
reduces to
 $$
\matr{\sigma}= C_4 \matr{\vep} + A_4 \matr{\dot{\vep}} \; ,
 $$
which is a simplest viscoelastic approximation at lowest
frequencies, or shear rates, of a general linear-response
expression $\sigma(t) = \int G(t-t') \dot{\vep}(t') dt'$. Another
known limiting case is that of an uncrosslinked nematic liquid
crystal. In this case all equilibrium (zero-frequency) elastic
moduli $C_{\rm i} $ and $D_{\rm i}$ are zero and the only
contribution to the stress and torque balance are the
Leslie-Ericksen eqs.~(\ref{sigma2}). In a low molecular weight
nematic, this is the full description of an anisotropic Newtonian
liquid. In a polymer nematic, again, one expects a complex
viscoelastic response function,  $G(t-t')$, with several
characteristic time scales, from the shortest Rouse time, to the
entanglement and diffusion times (if applicable) \cite{DoiEd}.
Equations~(\ref{sigma2}) are thus the low-frequency limit of such
complex nematic viscoelasticity.

The theory of elastomer nematohydrodynamics has 5 phenomenological
viscous coefficients. In a simplified single-viscosity model
\cite{letter} we have found remarkable effects the internal
director relaxation has on the macroscopic dynamic-mechanical
response of a nematic rubber. Here we follow the example of
ordinary liquid crystals, where both the simplified and the full
Leslie-Ericksen formalism have been used successfully over the
years.  In the practical calculation implementing the above ideas,
we now move from the one-constant analysis of \cite{letter} to a
full analysis.

\section{Shear stress and its relaxation}

\subsection{Simple shear deformations}\label{simple}
\begin{figure}
\centering
%\resizebox{0.37\textwidth}{!}{\includegraphics{shears.eps}}
\resizebox{0.37\textwidth}{!}{\includegraphics{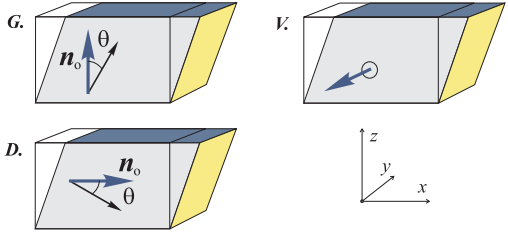}}
\caption[]{The geometry of simple shear experiment with three
principal orientations of the nematic director $\bm{n}_0$,
labelled $G$ (for $\bm{n}_0$ along the shear gradient), $D$
(displacement) and $V$ (vorticity). The small-amplitude shear
$\varepsilon_{xz} \sim \frac{1}{2} \varepsilon e^{{\rm i} \,
\omega t}$ is applied to the elastomer and the measured stress
$\sigma(\omega)$ provides the linear response modulus in each of
the three configurations. } \label{shears}
\end{figure}

In a study of linear response, we shall first examine three
principal simple shear geometries, as shown in Fig.~\ref{shears}.
These are also the geometries that one achieves in a typical
dynamic-mechanical experiment \cite{letter}. A geometry of
uniaxial extension, more commonly found in studies of equilibrium
stress-strain in elastomers, is less appropriate for an
oscillating regime because of possible slow relaxation
\cite{relax} and incomplete sample recovery on each cycle. The
simple shear $\varepsilon(t)$, assumed externally applied to the
sample (Fig.~\ref{shears}), is the single $xz$-component of the
full Cauchy strain, the same for each director setup $\bm{G}, \,
\bm{D}$ and $\bm{V}$. It also automatically satisfies the
necessary incompressibility constraint. We have then the symmetric
part $\varepsilon_{zx}^{(s)}=\frac{1}{2}\varepsilon$ and the
antisymmetric (body rotation) part expressed by
$\Omega_y=\frac{1}{2}\varepsilon$.

Note that the three principal shear geometries in
Fig.~\ref{shears} are the same as in the classical setting for
Miesovicz viscosity experiments. That is, if the director is kept
immobile (e.g. by a strong external field, which incidentally
would be hardly possible in elastomeric network under strain),
then the orientations $\bm{V}, \, \bm{D}$ and $\bm{G}$ correspond
to the Miesovicz viscosities $\eta_{\rm a}$, $\eta_{\rm b}$ and
$\eta_{\rm c}$, respectively. Of course, the whole point of the
present paper is to examine the effect the director freedom to
rotate is having on the viscoelastic response.

We have argued above that Frank effects are subordinate in the
overall balance of torques. Perhaps, this approximation needs to
be reconsidered if a sheared sample is very thin in the
$z$-direction: if the director is anchored on top and bottom
surfaces (which are the plates in a shear-sandwich experiment),
one expects a Frank energy density of the order $~K/d^2$. The
elastic energy density is set by $\mu$ and the two scales take on
equal importance when $d = \xi = \sqrt{K/\mu}$, the nematic
penetration depth.  Taking typical values $K \sim 10^{-11} {\rm
J/m^3}$ and $\mu \sim 10^5{J/m^3}$ one has $\xi \sim 10^{-8}{\rm
m}$.  Thus  in  practical situations with a sheared sample
thickness $d\sim 100 \, \mu$m and more, ignoring the effect
nematic director gradients seems to be a safe first approximation.

The remaining free energy density, eq.~(\ref{mainF}), takes the
form, in the three cases of Fig.~\ref{shears}:
\begin{eqnarray}
F_{\rm G} &=&\left(C_5+
\textstyle{\frac{1}{8}}[D_1-2D_2]\right)\vep^2
-\textstyle{\frac{1}{2}}(D_1-D_2) \vep \, \theta
+\textstyle{\frac{1}{2}}D_1 \theta^2 \nonumber  \\ F_{\rm D}
&=&\left(C_5+ \textstyle{\frac{1}{8}}[D_1+2D_2]\right)\vep^2
-\textstyle{\frac{1}{2}}(D_1+D_2) \vep \, \theta
+\textstyle{\frac{1}{2}}D_1 \theta^2 \nonumber \\ F_{\rm V}&=& C_4
\, \vep^2  \ ,  \label{ABC}
\end{eqnarray}
where the small change in director orientation, $\delta \bm{n}$,
is taken equal to the angle $\theta$. Clearly, one does not expect
director rotation to occur in the ``log-rolling'' geometry
$\bm{V}$.

The bilinear term $\sim \varepsilon \, \theta$ clearly reduces the
energy: if a strain $\varepsilon$ is imposed then the director
responds by adjusting $\theta$ to lower energy.  Given sufficient
time to reach equilibrium, $\theta$ adopts its optimal value for a
given deformation $\vep$. Returning this minimised value
$\theta_{\rm G,D}$ in (\ref{ABCoptimal}) to (\ref{ABC}), the free
energy at a given strain is also optimal:
\begin{eqnarray}
\theta_{\rm G,D} =  \frac{D_1 \mp D_2}{2 D_1} \, \varepsilon \ ,
\quad F_{\rm G,D} &\rightarrow& \left(C_5-\frac{D_2^2}{8D_1}
\right) \, \vep^2  \label{ABCoptimal}
\end{eqnarray}
Notice that if the nematic elastomer is ideal, then the free
energies in the cases $\bm{G}$ and $\bm{D}$ vanish
($C_5-D_2^2/8D_1=0$) because their geometry allows the director to
respond to the shear and internally relax.  Case $\bm{V}$ remains
elastically hard. In fact, because of the chosen restricted strain
geometry, Fig.~\ref{shears}, the response of even ideal elastomers
is actually quartic, rather than completely soft (true softness
requires some unconstrained extension as well as shear). The
molecular model \cite{review1,JMPS} yields the quartic penalty $
F= \frac{1}{2}\mu \ \frac{r^2}{(r-1)^2} \, \vep^4 $, which we are
neglecting in the present linear-response analysis.

The generalised force driving the director rotation angle
$\theta(t)$ to equilibrium is in fact a torque, $\partial
F/\partial \theta$. In $F$ it is the externally imposed shear
$\vep(t)$, with its time variation, that is the ultimate driving
agent. The flatness (softness) of the free energy $F(\vep)$ will
make the resulting dynamical response characteristically slow in
the geometries where director rotation is possible
\cite{teixeira}.  Alternatively, if the imposed dynamics, e.g.
$\varepsilon (t)$, is fast compared with that of $\theta(t)$, then
one will not attain the ideal states (\ref{ABCoptimal}) and the
moduli will become frequency-dependent and deviate, for dynamic
reasons, away from softness.  It is this rheological subtlety that
we now wish to examine.

\subsection{Viscoelastic response}
Driving a nematic elastomer by imposing a component of strain, in
general, leaves open the possibility of dynamical response of not
only the director but also the other components of strain (and
thus involving for instance the translational viscosities $A_1$,
$A_5$ and $A_5$).  Then there will be several coupled dynamical
equations for the system, see for instance \cite{teixeira} where
the response of a nematic elastomer to a step-extensional strain
is calculated. Here we confine ourselves to the dynamical response
to imposed simple shear with other components of strain clamped.
An elastic torque acting on nematic director $\bm{n}$ in certain
deformation geometries is resisted by a linear viscous torque.
Describing the director rotation by a small angle $\theta$, cf.
Fig.~\ref{shears}, and continuing to ignore inertial effects, the
balance of torques and forces is expressed by
\begin{equation}
\frac{\delta F}{\delta \theta} + \frac{\delta (T\dot{s})}{\delta
\dot{\theta}} =0 . \label{balance}
\end{equation}
The dynamical equation describing the evolution of the director
approaching its equilibrium is given, for the two geometries where
director rotation is present:
\begin{eqnarray}
\bm{(G:)} \ \ \ \ && \gamma_{\rm 1}\, \dot{\theta} = -D_1 \theta +
{\textstyle{\frac{1}{2}}}(D_1-D_2)\, \vep(t)  \nonumber \\ &&
\qquad \qquad + {\textstyle{\frac{1}{2}}}(\gamma_{\rm 1}-
\gamma_{\rm 2})\, \dot{\vep}(t) \label{ABCdot}
\\
 \bm{(D:)} \ \ \ \ &&\gamma_{\rm 1}\, \dot{\theta} = -D_1 \theta +
{\textstyle{\frac{1}{2}}}(D_1+D_2)\, \vep(t) \nonumber \\ &&
\qquad \qquad + {\textstyle{\frac{1}{2}}}(\gamma_{\rm 1}+
\gamma_{\rm 2})\, \dot{\vep}(t) \; . \nonumber
\end{eqnarray}
These linear, inhomogeneous differential equations are easily
solved. After the transient relaxation $\theta \simeq \theta_0
e^{-(D_1/\gamma_{\rm 1} ) t}$ associated with starting the strain
oscillations has completely relaxed, the steady-state response is
given by the particular solutions with $\vep \sim e^{{\rm i} \,
\omega t}$:
\begin{eqnarray}
\theta_{\rm G,D}(t) &=& \int_{-\infty}^t dt' \,
e^{-(D_1/\gamma_{\rm 1} [t-t'])} \label{ABCsols}  \\ && \times
\left( \frac{D_1 \mp D_2}{2\gamma_{\rm 1}}\, \vep(t') +
\frac{1}{2}(1 \mp  \gamma_{\rm 2}/\gamma_{\rm 1})\, \dot{\vep}(t')
\right) \; , \nonumber
\end{eqnarray}
where the signs $-$ or $+$ correspond to the $\bm{G}$ or $\bm{D}$
geometry, respectively. The solutions depend on a characteristic
time for director relaxation, $\tau_{\rm 1}=\gamma_{\rm 1} /D_1$,
or equivalently, the characteristic frequency of the response
scales like $D_1/\gamma_{\rm 1} $.  We can now give a more
concrete expression for the new dimensionless number ${\sf Ne} =
\gamma_1 \nabla v / (D_1 \theta)$, since $\theta \sim \varepsilon$
and $\nabla v \sim {\dot \varepsilon} \sim \omega \varepsilon$,
whence: \begin{eqnarray} {\sf Ne} = \gamma_1 \omega /D_1 = \tau_1
\omega \label{newnumber}\; .\end{eqnarray} Below we shall see that
all depends on $\tau_1 \omega$ -- when it is small, we have soft
or semi-soft equilibrium elastic response, when it is large, we
have hardening because director relaxation does not keep pace with
changing strain.

The linear ``nominal stress'' in response to the imposed simple
shear deformation $\vep(t)$ is given by the sum of elastic and
viscous stress functions $\sigma= \sigma_{\rm el} + \sigma'_{\rm
visc}$, that is
\begin{equation}
\sigma = \frac{\partial F}{\partial \vep} + \frac{\partial
(T\dot{s})}{\partial \dot{\vep}} .
\end{equation}
In the principal shear geometries, the stress expressions at a
given frequency of imposed strain take the form
\begin{eqnarray}
\sigma_{\rm G}(\omega)&=&2 \left(C_5+ \textstyle{\frac{1}{8}}
[D_1-2D_2] \right)\, \vep(\omega) - \textstyle{\frac{1}{2}}
(D_1-D_2) \, \theta(\omega) \nonumber
\\
&& +2 \left(A_5+ \textstyle{\frac{1}{8}} [\gamma_1-2\gamma_2]
\right)\, \dot{\vep} - \textstyle{\frac{1}{2}} (\gamma_1-\gamma_2)
\, \dot{\theta} \label{sG} \\
 \sigma_{\rm D}(\omega)&=&2 \left(C_5+ \textstyle{\frac{1}{8}}
[D_1+2D_2] \right)\, \vep(\omega) - \textstyle{\frac{1}{2}}
(D_1+D_2) \, \theta(\omega)  \nonumber \\ &&+ 2\left(A_5+
\textstyle{\frac{1}{8}} [\gamma_1+2\gamma_2] \right)\, \dot{\vep}
- \textstyle{\frac{1}{2}} (\gamma_1+\gamma_2) \, \dot{\theta}
\label{sD} \\
 \sigma_{\rm V}(\omega)&=& 2C_4 \, \vep(\omega) +
 2A_4 \, \dot{\vep} \; . \label{sV}
\end{eqnarray}
One finds two kinds of viscous stress terms, of different orders
of magnitude. The contribution $ \gamma  \dot{\theta} \sim \gamma
 \dot{\vep}$ is of the same order as the elastic terms, as
indicated by the torque balance eq.~(\ref{ABCdot}). In contrast,
the terms $\sim A_{4,5} \dot{\vep}$ are of the order $A \dot{\vep}
\sim C (\omega \tau_{\rm R}) \vep \ll C \vep$ at frequencies below
Rouse values. The linear viscoelastic theory we are considering is
applicable at much lower frequencies, where the most interesting
physics is due to director relaxation and mechanical softness.

The effective response modulus in the ``log-rolling'' geometry
$\bm{V}$ is unchanged by the nematic director dynamics,
$G(\omega)=2C_4+ 2{\rm i}\omega  A_4$ (the loss modulus
negligible, as discussed above). In two other geometries, where
the director rotation does take place, the dynamic modulus is
modified by the internal director relaxation. Substituting the
Fourier transforms of equations (\ref{ABCsols}) into the
expressions (\ref{sG}) and (\ref{sD}), we obtain the nominal
stress in the form $\sigma(\omega) = G(\omega) \vep(\omega)$.
Remarkably, although perhaps predictably, the response in these
two geometries is exactly the same: $\sigma_{\rm G} = \sigma_{\rm
D}$, despite the difference in the rotations $\theta_{\rm G}$ and
$\theta_{\rm D}$. The corresponding storage and loss moduli are
given by the real and imaginary parts of the effective complex
modulus $G(\omega)$. It has a single-relaxation time behaviour
with a characteristic frequency $\omega_{\rm 1} = D_1/\gamma_{\rm
1}$ or the corresponding relaxation time $\tau_{\rm 1} =
1/\omega_{\rm 1}$. The dynamic moduli, in both $\bm{G}$ and
$\bm{D}$ geometries, can then be written in a universal form:
\begin{eqnarray}
G'(\omega) &=& 2(C_5 - D_2^2/8D_1) \label{G1universal}  \\ && +
\frac{\left(\omega \tau_1 \right)^2}{1+ \left(\omega \tau_1
\right)^2} \frac{\left(D_2 \gamma_1 - D_1 \gamma_2  \right)^2}{4
D_1 \gamma_1^2} \nonumber
\\
 G''(\omega) &=& \frac{\omega \tau_1}{1+ \left(\omega \tau_1 \right)^2}
\frac{\left(D_2 \gamma_1 - D_1 \gamma_2  \right)^2}{4 D_1
\gamma_1^2} \label{G2universal} \\ &+& \omega \tau_1 \left(
{\textstyle{\frac{1}{2}}} D_1(\gamma_2/\gamma_1)^2 - 2A_5/\tau_1
\right) \; . \nonumber
\end{eqnarray}

\begin{figure}
\centering
%\resizebox{0.45\textwidth}{!}{\includegraphics{modO.eps}}
\resizebox{0.45\textwidth}{!}{\includegraphics{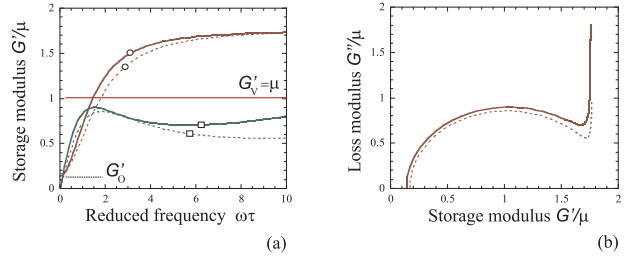}}
\caption[]{(a) The frequency dependence of storage ($G'$, circles)
and loss ($G''$, squares) moduli in the two shear geometries
$\bm{G}$ and $\bm{D}$ in units of rubber modulus $\mu$. There is
no loss in the $\bm{V}$-geometry and, since $G'_{\rm V}=\mu$, the
plot represents the variation of the ratios $G'_{\rm G}/G'_{\rm
V}$ and $G'_{\rm D}/G'_{\rm V}$. At $\omega \rightarrow 0$ both
$G' \rightarrow 2C_5^R$. \ (b) The Cole-Cole plot, a variation of
$G''$ with $G'$. A semicircular shape indicates a single
relaxation time response. In both plots, solid lines are for
$Q=0.58$, dashed lines for $Q=0.35$. } \label{dynresp}
\end{figure}

Figure~\ref{dynresp} shows the example of frequency dependence of
eqs.~(\ref{G1universal}) and (\ref{G2universal}). In order to plot
these functions, we need to make several numerical assumptions
about the parameters. We thus take,  arbitrarily for illustration
purposes, $r=1+1.5Q$ at a fixed temperature with  values of the
nematic order parameter corresponding to  deep in a nematic phase,
$Q =0.58$ (so that the chains are weakly prolate, $r\approx
1.87$), or corresponding to near a transition point, $Q=0.35$. The
ratio $\gamma_{\rm 2}/\gamma_{\rm 1} \approx 1$ and the viscous
constant $A_5 \approx 0.1 \tau_1 C_5$. Finally, the semi-soft
addition to the coupling constant $D_1$ (see the Appendix) is
taken as $a_1=0.1$. For these values, plots of $G'$ and $G''$
reveal the expected single-relaxation behaviour. At zero frequency
of imposed strain oscillations the real (storage) modulus  in both
shear geometries $\bm{G}$ and $\bm{D}$ is equal to
 $$G'_{\rm o} \equiv 2 C_5^{\rm R} = 2(C_5 - D_2^2/8D_1)\; . $$
 The measure of non-vanishing renormalised shear
modulus $C_5^{\rm R}\neq 0$ is the characteristic parameter of
semi-softness in non-ideal nematic elastomers. The high frequency
response for $\omega \tau_{\rm 1} \gg 1$ is
 \begin{eqnarray} G'_{\infty}= 2C_5
- {\textstyle{\frac{1}{2}}}D_2 (\gamma_{\rm 2}/\gamma_{\rm 1}) +
{\textstyle{\frac{1}{4}}} D_1 (\gamma_{\rm 2}/\gamma_{\rm
1})^2.\label{highf}
\end{eqnarray}
The characteristic single-relaxation time behaviour seen in
fig.~\ref{dynresp} is due to our assumption about separation of
time scales, $\tau_1 \gg \tau_{\rm R}$. For that reason, we are
able to distinguish the plateau in $G'$, the maximum and the
``high-frequency'' drop in $G''$ and a semicircular shape of
Cole-Cole plot \cite{ferry}. For the chosen value of $A_5$, one
can begin to see the next rise in $G''$, when the frequency
approaches the next characteristic point, the Rouse frequency,
leading towards the glass plateau value $G'  \sim 10^{11} \,
\hbox{Pa}$ at high frequency. That transition would correspond to
\underline{another} semi-circular Cole-Cole plot at much higher
values of $G'$; we can see the beginning of that graph segment in
fig.~\ref{dynresp}(b). With the assumed separation of nematic and
rubber relaxation time scales (and the assumed ideal polymer
network, with no entanglements and free dangling ends, which are
known to lead to slow relaxation), the mechanical losses at low
frequencies are only through the lag in director rotation.

The initial expressions, eqs.~(\ref{sG}) and (\ref{sD}), differ in
sign of terms with $D_2$ and $\gamma_2$ constants. This reflects
the tendency to align or repel the director by these terms, the
roles being reversed when we interchange prolate and oblate
symmetry ($r > 1$ and $r < 1$, the sign of both $D_2$ and
$\gamma_2$ reversing on this exchange). However, when the director
relaxation (\ref{ABCsols}) is substituted back into the stress
expressions, one only finds the products $D_2 \gamma_2$, and a
corresponding unique effective complex modulus. Another feature of
$G^*(\omega)$ is the non-dimensional ratio $\gamma_{\rm
2}/\gamma_{\rm 1}$, which is a familiar parameter in the dynamics
of ordinary liquid crystals \cite{degenbook} in the context of
director flow alignment or tumbling. As the brief discussion of
modified torque balance in Section~3 indicates, one does not
expect a steady-state tumbling in elastomers, where there is
memory of the original director through the elastic energy.
Tumbling and analogous effects may only be found in transient
regimes or in non-linear elasticity, to which we return elsewhere.

\begin{figure}
\centering
%\resizebox{0.45\textwidth}{!}{\includegraphics{modT.eps} }
\resizebox{0.45\textwidth}{!}{\includegraphics{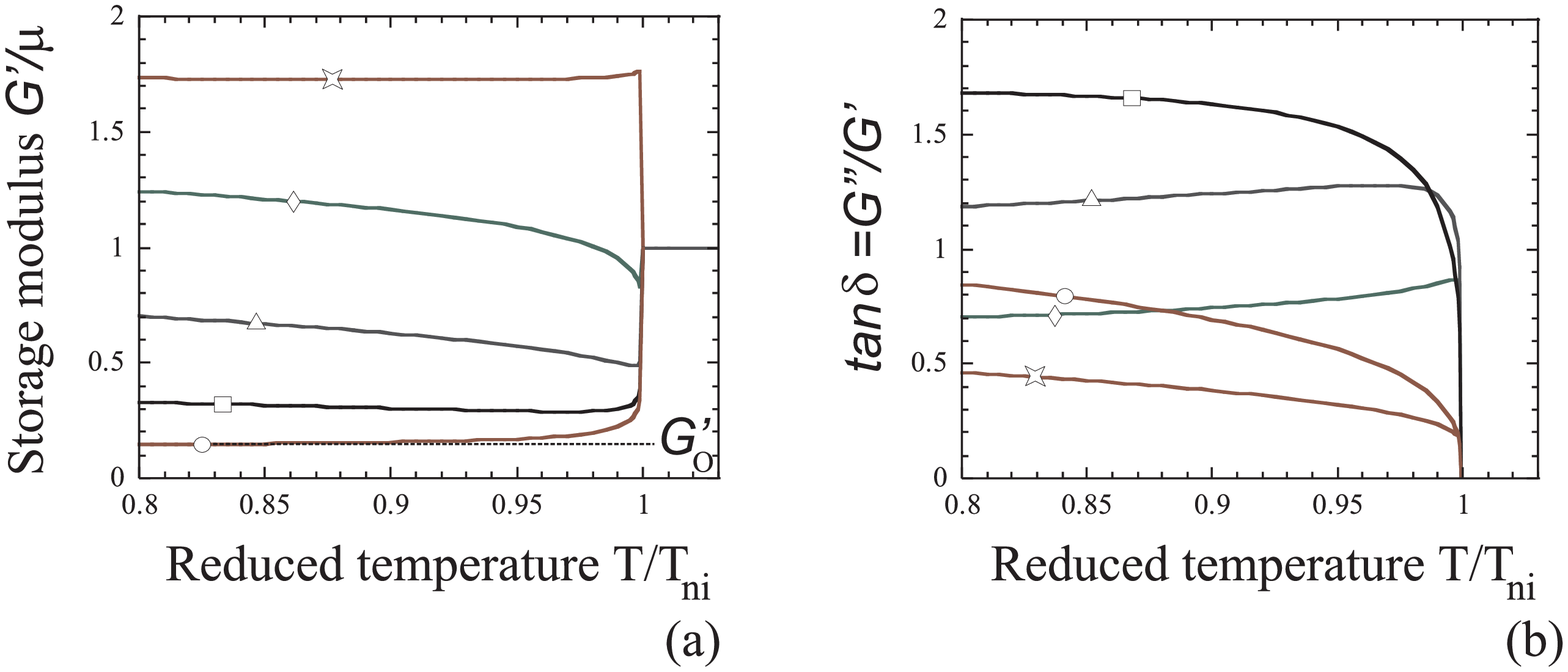} }
\caption{(a) Plot of reduced storage moduli $G'/\mu$ against
reduced temperature for a number of increasing reduced
frequencies:  $\tilde{\omega} =0.1$ (circles), $0.5$ (squares),
$1$ (triangles), $2$ (diamonds) and $10$ (stars). Moduli approach
the semi-soft constant $G'_{\rm o}$ at zero frequency and low
temperature, cf. Fig.~\ref{dynresp}(a). An apparent critical
behaviour near the transition point $T_{\rm ni}$ is due to the
nematic order parameter variation $Q(T)$ appearing in the elastic
and viscous coefficients. \ (b) Plot of loss factor $\tan \,
\delta = G''/G'$ against reduced temperature for a number of
increasing frequencies. The curves are labeled the same as in
(a).} \label{modT}
\end{figure}

Variation of dynamic complex moduli with temperature is very
interesting to examine graphically, as well as analytically.
Figure~\ref{modT}(a) shows the dependence of storage modulus $G'$
on the reduced temperature $T/T_{\rm ni}$, for a sequence of
increasing frequency values. In order to produce these plots, we
must take a further assumption about the variation of the order
parameter $Q(T)$. We take, rather arbitrarily, $Q \approx
(1-T/T_{\rm ni})^{0.33}$ simply because this was an approximate
fit to the experimental measurement of $Q(T)$ in \cite{letter}. We
also take, following  classical nematics, $\gamma_{\rm 1} = g_1
Q^2$ and $\gamma_{\rm 2}= g_2 Q$ and further fix $g_2/g_1 =
\frac{1}{2}$ to have $\gamma_{\rm 2} \approx \gamma_{\rm 1}$ at
$Q\sim 0.5$. These rather limiting assumptions nevertheless
usefully illustrate the qualitative behaviour of the linear
elastic response functions both deep in the nematic phase and near
the assumed ``critical point" $T_{\rm ni}$.

The characteristic time of director relaxation $\tau_1
=\gamma_{\rm 1}/D_1$ is a function of temperature through its
dependence on nematic order parameter. Both the coupling constant
$D_1$ and the rotational viscous coefficient $\gamma_{\rm 1}$ are
functions of $Q(T)$, as discussed in Section~3. Hence the
relaxation time $\tau_1$ should have a weak $Q$-dependence in the
nematic phase, which should change to $\tau_{\rm 1} \sim |Q |$ in
the vicinity of the nematic-isotropic transition point.  We
discuss subtle limit problems and the form of $G(T,\omega)$ as a
function of $T$ for various fixed $\omega$ in the Appendix. The
combination $\omega \tau_1$ can be rewritten as
 $$\omega \tau_1
\equiv \omega \frac{\gamma_{\rm 1}}{D_1} =  \omega \left(
\frac{g_1}{\mu} \right) \frac{Q^2}{a_1|Q |+(r-1)^2/r},
 $$
where we recall that $r \sim 1 + \beta Q$ (in practice, the linear
relation holds to high values of $Q$ \cite{letter,fgw}). We shall
continue taking $\beta = 1.5$ in our illustration. One can also
scale $\omega$ by $\mu/g_1$ to give a dimensionless frequency
$\tilde{\omega}$; thus $\omega \tau_1 =\tilde{\omega} f(Q)$ where
the non-dimensional function $f(Q)$ is clearly $f \sim |Q |$ as $Q
\rightarrow 0$.

The cusps near $T_{\rm ni}$ seen in Fig.~\ref{modT}(a) are a
consequence of our assumption of the critical form $Q(T) \sim
(1-T/T_{\rm ni})^{.33}$. For a real, non-ideal monodomain nematic
elastomer with of necessity a high-temperature paranematic state
and a residual small $Q$ above $T_{\rm ni}$, the cusps are rounded
off but the general form survives, including the non-monotonic
variation $G'(T)$ at high frequencies. We note in passing, that
the experimental result for $G'(T)$ \cite{letter} is very close to
these predictions: although the very high frequency was not
achieved, the gradual departure of $G'(T)/\mu$ from the universal
low-frequency curve was clearly registered.

Figure \ref{modT}(b) also shows the unusually high loss tangent
arising from the real and imaginary parts of complex modulus. The
new loss mechanism, due to internal director relaxation, is
additional to all classical losses occurring in polymer systems --
in particular those occurring near the glass transition. The
information about the latter is, as before, contained in the
essential frequency and temperature dependence of rubber-elastic
constant $\mu$, of which the expressions for $F$ and $T \dot{s}$
are the low-frequency limit. There is a striking contrast between
the fig.~\ref{modT}(b) with unusually high values of $\tan \,
\delta$ over a broad temperature range and the traditional
distinct loss peaks in classical polymers and rubbers. Physically,
the maximal loss occurs where the the imposed strain frequency,
$\omega$, matches the director relaxation rate $1/\tau_1$. We have
a crossover from the director keeping up with strain and therefore
allowing the the system to be very soft, to where the director
fails to respond quickly enough and the system becomes hard. The
large change of modulus when one crosses this frequency interval
is manifested, by the Kramers-Kr\"onig relations, in a very high
loss peak.

\section{Conclusions}

We have developed a model of time-dependent response and rheology
of nematic elastomers by coupling nematic rubber elasticity with
nematohydrodynamics. The theory represents the low-frequency limit
of general viscoelastic description of nematic polymer networks
(which is not known). By making use of significant separation
between the natural nematic relaxation time scale and that of a
rubber network, we examine the dissipation at relatively low
frequencies due to internal director relaxation in a deformed
elastomer. Because of the lagging of director rotation in response
to time-dependent imposed shears, soft or semi-soft response can
only be partially achieved, the hardening depending on frequency
and on temperature. The nemato-elastic-hydrodynamic equations are
derived in generality and solved for the rheology of a particular,
important example, that of simple oscillating shear, giving rise
to unusual temperature dependence of response moduli and loss
tangents ($\tan\delta \sim 1.5 - 2$).  Thus these results will be
important for a wide variety of applications in damping,
acoustics, frequency and directional dependent materials, and so
on.

We have already reported on rheological experiments \cite{letter}
that show the qualitative features of our theory.  These and light
scattering experiments \cite{stille,Meyer_light} give values of
Frank constants, rubber constants $C_5$, $D_1$, $D_2$ and the
rotational viscosity coefficient $\gamma_1$ for nematic elastomers
(and the $\gamma_1$ also for the corresponding polymer melts). The
$C$'s and $D$'s can also be deduced from several static
measurements, in all cases giving typical rubber values $\sim
10^3-10^5$~Pa, modulated by the discussed $Q$-dependence. Frank
constants take values typical of classical nematic liquids, $K
\sim 10^{-11}$~N. But the $\gamma$ values are enhanced by
polymerisation. This is not surprising since on reorienting a
chain one must translate it bodily.  An extended object sweeps
through space and the dynamics are complex. In a network this is
even less straightforward because the chain is mechanically tied
to a matrix and must resolve its motions with those of other
chains. It turns out that the $\gamma$'s are further enhanced,
reaching values of order $\sim 10^3$~Pa.s and more, leading to the
mentioned separation of characteristic time scales.\\

\noindent We appreciate valuable discussions with S.M Clarke, H.
Finkelmann,  T.C. Lubensky, R.B. Meyer, P.D. Olmsted and  W.
Stille.

\appendix
\section{Semi-softness and rheology}

At fixed temperature in the nematic state, Fig.~\ref{dynresp}
shows the rheological consequences of nematohydrodynamics in
elastomers -- as frequency drops and director relaxation had keep
pace with imposed strains, the response gets ever softer.  In the
limit $\omega \tau_1 \rightarrow 0$ it either vanishes (ideal
softness) or becomes very small (semi-softness) as has been
confirmed by many quasi-static experiments.  We wish to comment on
the mechanisms for softness and non-ideality and then how, at
various fixed frequencies, one can see how $G'$ tends to
conventional rubber elastic response as $T$ increases and
$Q\rightarrow 0$.

Soft shear response occurs because anisotropic network chains
accommodate a macroscopic shape change by rotating their
distribution of shapes at constant entropy.  When chains tend to
isotropy, $Q \rightarrow 0$ and $r \rightarrow 1$, shape change
can only be accommodated by chains distorting their distribution,
thereby decreasing their entropy and increasing their free energy.
The modulus is then $\mu$.  We must see how this classical limit
is achieved; after all the cancellation $C^R_5 = C_5 - D_2^2/8D_1
= 0$ appears to hold for all $r$ including $r\rightarrow 1$.

Semi-softness is expected when there are fluctuations of
composition, rod-like crosslinks, sources of random nematic field,
and any other form of non-ideality that prohibits the finding of
an isotropic reference state \cite{JMPS,GL}.  Then there are
additions to the Trace formula (\ref{F0}) which are of the form $a
\sin^2\theta$ for simple shears, see for instance
\cite{TWV_stripes}.  The degree of semi-softness, $a$, can be
calculated directly from various models of non-ideality, or
measured.  For continuum theory such additions appear in the
eqs.~(\ref{params}) for the $C_i$ and $D_i$, and upset the
cancellations in (\ref{renorm}), yielding a $C^R_5 \ne 0$, which
is the non zero limit at $\omega\tau_1 \rightarrow 0$ in
Fig.~\ref{dynresp} for $G'$. However, the additions inspired by
$a$ clearly must vanish as $Q \rightarrow 0$ and $r \rightarrow 1$
and the delicate limit question must still be resolved.  In
particular in (\ref{renorm}) $C^R_5 \rightarrow C_5 \rightarrow
\frac{1}{2} \mu$ as $ r \rightarrow 1$.

In the ideal case there is no high temperature order, softness is
perfect and for $T < T_{\rm ni}$ the modulus $C^R_5$ is
identically zero.  Several systems, depending on their
thermomechanical history, are close to this \cite{JMPS}.  Since
ideally $Q$ jumps to zero at $T = T_{\rm ni}$, there is no limit
problem.  $D_1$ and $D_2$ cease to exist, the analysis
(\ref{ABCoptimal}) is invalid and $C_5$ is not renormalised.

In the semi-soft case, (\ref{renorm}) places bounds on the form
the non ideal additions to $D_1$ can take.  Since $D_1 \sim Q^2$
and $D_2 \sim Q$, in the ideal case the limit $D_2^2/D_1$ is
finite as $Q \rightarrow 0$.  If however $D_1$ has additions $D_1
\sim (Q^2 + a_1 Q)$, then the limit of  $D_2^2/D_1$ vanishes as $Q
\rightarrow 0$, eliminating any renormalisation.  Explicit
calculations agree with symmetry arguments that the additions to
$D_1$ are indeed $\sim Q$ and reflect the thermomechanical history
of the material. For instance fluctuations (due to compositional
fluctuations in the polymers that make up the network) in the
effective order felt by a chain, $Q_{\rm eff} = (1 + \delta)Q$
(with $\langle \delta \rangle = 0$), yield for the modulus:
\begin{eqnarray}
D_1  = \mu ' \left[ \frac{(r-1)^2}{r} + a_1 Q\right]
\label{nonsoft}
\end{eqnarray} where $\mu' =
\left(\frac{\det{\matr{\ell}_f}}{\det{\matr{\ell}}}\right)\, \mu$
and
 $$ a_1 = 3 \langle \delta^2 \rangle \frac{Q_f (3 +
2Q_f)}{(1+2Q_f)(1-Q_f)}.
 $$
Here formation conditions for the network are denoted by a
subscript $f$: the step length tensor at formation is
$\matr{\ell}_f$ with order parameter $Q_f$.  The
determinant-factors reflect spontaneous shape changes since
formation which can be very large (up to several hundred \%).  The
sign of the coefficient $a_1$ is that of $Q_f$.  It is the order
at network formation that induces the residual order $Q$ at high
temperatures and the sign of that order follows that of the
formation order.  Thus the combination $a_1Q$ is always positive
and thus the non-ideal additions to $D_1$ in (\ref{nonsoft}) are
really of the form $| Q |$, which we have used in section
\ref{symmetries}.

Thus at sufficiently low frequencies one observes a rather small
modulus $G' = C^R_5 \ge 0$, which rises with increasing
temperature to reach the classical value $\mu$ at $Q\rightarrow 0$
at $T_{\rm ni}$. If, at fixed frequency, on dropping the
temperature the combination $\omega \tau_1$ becomes appreciable
(due to increasing $\tau_1$), the modulus $G'(T)$ will depart from
the universal low-frequency form $G'_{\omega\tau_1 \rightarrow
0}$, see Fig.~\ref{modT}.

\end{document}